\documentstyle[12pt]{article}
\newcommand{\NP}[1]{Nucl. \ Phys.}
\newcommand{\PL}[1]{Phys. \ Lett.}
\newcommand{\p}[1]{\partial}
\newcommand{\PRL}[1]{Phys.\ Rev.\ Lett. }
\newcommand{\AP}[1]{Ann.\ Phys. }
\newcommand{\CMP}[1]{Commun.\ Math.\ Phys. }
\newcommand{\TMF}[1]{Theor.\ Math.\ Phys. }
\newcommand{\CQG}[1]{Class.\ Quant.\ Grav. }
\newcommand{\MPL}[1] { Mod. Phys. Lett. }
\newcommand{\IJMP}[1] { Int. J. Mod. Phys. }
\begin{document}

\title{
Divergences of Discrete States Amplitudes
and Effective Lagrangian in 2D String Theory.}
\author{ I.Ya.Aref'eva \thanks{ Steklov Mathematical Institute,
Vavilov 42, GSP-1, 117966, Moscow, Russia} \\
 and\\
 A.P.Zubarev \thanks
{Supported in part by Moscow Physical Society
Grant}}
\date {May, 1992}

\maketitle
\begin{abstract}
Scattering amplitudes for discrete states in 2D string theory are considered.
Pole  divergences of tree-level amplitudes are extracted and  residues are
interpreted as renormalized amplitudes  for discrete states. An effective
Lagrangian generating renormalized amplitudes for open string is written and
corresponding Ward identities are presented. A relation of this Lagrangian
with  homotopy Lie algebra is discussed.
$$~$$
$$~$$
$$~$$
$$~$$
$$~$$
$$~$$
$$~~~~~~~~~~~~~~~~~~~~~~~~~~~~~~~~~~~~~~~~~~~~~~~~~~~~{\bf SMI-2-92}~$$
\end{abstract}

\newpage
%%%%%%%%%%%%%%%%%%%%%%%%%%%%%%%%%%%%%%%%%%%%%%%%%%%*****************
\section   {Introduction}
%*********************************************************************

Recently Witten and Zwiebach  \cite {WZ} and E.Verlinde  \cite {V} have
considered Ward identities for discrete states  amplitudes
\cite {Gross}- \cite {IO}
in 2D string theory.
These identities express a symmetry which is important in understanding of
background independent formulation of the corresponding string field theory.
It was noted  \cite {WZ} a relation of these Ward identities with homotopy
Lie algebra \cite {St}.

In this note we consider these amplitudes in more details.
In fact correlation functions for discrete states diverge because external
momentum coincide with poles of amplitudes. We use an analytic regularization
and interpret a residue with respect to regularization parameter $\epsilon$ as
renormalized correlation functions for discrete states. These
correlation functions satisfy Ward identities. Such renormalization
corresponds to renormalization of coupling constant
and wave function in string field theory  \cite {AZ} and it
should correspond to
renormalization of the Verlinde master equation  \cite {V}.

We present an effective Lagrangian generating renormalized correlation
function.
It turns out that this Lagrangian is related with the structure of the homotopy
Lie algebra.

%*********************************************************************
\section   {Tree level N-point amplitudes}
%*********************************************************************
Below we carry out the calculations of the simplest on-shell amplitudes for
discrete states. From this results it is not hard to infer the general
structure.

Let us consider the tree level 4-point open string amplitude for discrete
states.
Assume that  three of them are (+) states
\begin {equation} %---------------------------------------------------
                                                          \label {1}
Y^{+}_{s_{i},n_{i}}=cW^{+}_{s_{i},n_{i}}
\end   {equation} %---------------------------------------------------
$i=1,2,3$, and the fourth  is (-) state
\begin {equation} %---------------------------------------------------
                                                          \label {2}
Y^{-}_{s_{i},n_{i}}=cW^{-}_{s_{i},n_{i}}
\end   {equation} %---------------------------------------------------
$i=4$, where
\begin {equation} %---------------------------------------------------
                                                          \label {3}
W^{\pm}_{s,s-n}=\sqrt{(2s-n)!/n!(2s)!}
\underbrace {[H^{-},...[H^{-}}_{n},W^{\pm}_{s,s}]...],~~
\end   {equation} %---------------------------------------------------
$$H^{-}=\oint \frac{dz}{2\pi i }e^{-i\sqrt {2}X}(z),
{}~~W^{\pm}_{s,s}=e^{i\sqrt{2}sX}e^{\sqrt{2}(1\mp s)\phi},$$
$s$ is positive integer or half integer.

In this case we have
\begin {equation} %---------------------------------------------------
                                                          \label {4}
{\cal A}_{4}((s_{1},n_{1})^{+}(s_{2},n_{2})^{+}(s_{3},n_{3})^{+}
(s_{4},n_{4})^{-})=
\end   {equation}%---------------------------------------------------
$$\int <Y^{+}_{s_{1},n_{1}}(z_{1})Y^{+}_{s_{2},n_{2}}(z_{2})
Y^{+}_{s_{3},n_{3}}(z_{3})
\oint b(z)dzY^{-}_{s_{4},n_{4}}(z_4)>d\mu (z_{3}).
$$
These amplitudes contain divergences.
For example, the amplitude for four tachyon states with exceptional momentum
is ill-defined,
$${\cal A}_{4}((1,1)^{+}(1,-1)^{+}(1,1)^{+}(1,-1)^{-})=$$
\begin {equation} %------------------------------------------------
\label {5}
=\int dz<(Y^{+}_{1,1})(\infty)(Y^{+}_{1,-1})(1)
(W^{+}_{1,1})(z)(Y^{-}_{1,-1})(0)>=
\int _{0}^{1}dz \frac{1}{z^{2}(1-z)^{2}}.
\end   {equation} %---------------------------------------------------
The integral  is divergent and one should use some regularization procedure.
Results will depend on the regularization. We use a special
analytical regularization and corresponding integral
understand in sense of the analytical continuation
\begin {equation} %---------------------------------------------------
                                                          \label {9}
reg_{\epsilon}(\int _{0}^{1}dzz^{-n}(1-z)^{-m})=
\int _{0}^{1}dzz^{-n-\epsilon}(1-z)^{-m-\epsilon}=
B(1-n-\epsilon ,1-m-\epsilon),
\end   {equation} %---------------------------------------------------
 Note that just with such analytical continuation
on external momentum one deals in the usual D=26 critical bosonic string.
Under this convention the amplitude (\ref {5}) has a pole
with respect to the regularization
parameter
\begin {equation} %---------------------------------------------------
                                                          \label {5a}
{\cal A}_{4}((1,1)^{+}(1,-1)^{+}(1,1)^{+}(1,-1)^{-})
\sim \frac{2}{\epsilon}
\end   {equation} %---------------------------------------------------

Some amplitudes (\ref {4}) under  regularization (\ref {9}) have not
singularities. For example, the amplitude for four vector states
is finite
$${\cal A}_{4}((1,0)^{+}(1,0)^{+}(1,0)^{+}(1,0)^{-})=$$
\begin {equation} %---------------------------------------------------
\label {5'}
=\int _{0}^{1}<(c\partial x)(\infty)(c\partial x)(1)
(\partial x)(z)(c\partial x)(0) e^{2\sqrt{2}\phi }> =\int _{0}^{1}dz
[\frac{1}{z^{2}}+\frac{1}{(1-z)^2}+1]=-1.
\end   {equation} %---------------------------------------------------
It is easy to see that the leading singularities on $\epsilon$ can be
extracted by inserting the sum over cohomology classes with ghost number 1
(physical states) in  the correlation functions (\ref {4}). Indeed, to
calculate
the  s-channel amplitude (\ref {4}) let us perform the operator product
expansion
of operators $W^{+}_{s_{3},n_{3}}(z)
$ and $Y^{-}_{s_{4},n_{4}}(0)$,
\begin {equation} %---------------------------------------------------
                                                          \label {6'}
W^{+}_{s_{3},n_{3}}(z)Y^{-}_{s_{4},n_{4}}(0)=\sum _{i}\frac{1}{z^{i}}
{\cal O}_{i}(0).
\end   {equation} %---------------------------------------------------
Physical states $Y^{-}_{s',n'}$
(non-trivial cohomology classes)
appeared in ${\cal O}_1$ due to the OPE of operators $W^{+}$ and $W^{-}$,
see  \cite {KP},  give leading singularities (l.s.)
and we can write
\begin{equation} %---------------------------------------------------
            \label {6}
{\cal A}^{l.s.}_{4}=\sum _{s',n'}\int _{0}^{1}
\frac{dz}{z^{1+\epsilon}}
<Y^{+}_{s_{1},n_{1}}(\infty ) Y^{+}_{s_{2},n_{2}}(1)Y^{-}_{s',-n'}(0)>
<Y^{+}_{s',n'}(\infty )Y^{+}_{s_{3},n_{3}}(z)Y^{-}_{s_{4},n_{4}}(0)>
\end{equation} %---------------------------------------------------

Therefore we have shown that the 4-point amplitude under the special analytical
regularization has the pole form
\begin {equation} %---------------------------------------------------
                                                          \label {11}
{\cal A}_{4}= \frac{\tilde{{\cal A}_{4}}}{\epsilon}+O(1).
\end   {equation} %---------------------------------------------------
Just this function $\tilde{{\cal A}_{4}}$ we want interpret as the 4-point
correlation function for discrete states.
The physical meaning of the above calculations is clear,
since we do nothing but extract the  residue of $s$-channel pole.

Moreover, $\tilde{{\cal A}}_4=Res {\cal A}_4$
can be expressed in terms of 3-point functions (\ref {6}).
3-point functions have been considered in  \cite {KP} and have the form

\begin {equation} %---------------------------------------------------
                                                          \label {10}
<Y^{+}_{s_{1},n_{1}}Y^{+}_{s_{2},n_{2}}Y^{-}_{s_{3},n_{3}}>
=f_{s_{1}n_{1},s_{2}n_{2}}^{s_{3}n_{3}}
\end   {equation} %---------------------------------------------------
$$f_{s_{1}n_{1},s_{2}n_{2}}^{s_{3}n_{3}}
=\delta _{s_{1}+s_{2}-1,s_{3}}\delta
_{n_{1}+n_{2},-n_{3}} \tilde
{f}_{s_{1}n_{1},s_{2}n_{2}}. $$
For  integers $s_{1}$ and $s_{2}$ one has
$\tilde{f}_{s_{1}n_{1},s_{2}n_{2}}=s_{2}n_{1}-s_{1}n_{2}$.
In (\ref {10}) the following changing of
the normalization of the discrete states
$~~Y^{+}_{s,n}$ $~\to~$ $1/\tilde{N}(s,n)~Y^{+}_{s,n}~,~~~$
$\tilde{N}(s,n)=-\sqrt{s/2}(2s-1)!N(s,n),$
 $~N(s,n)=$ $\sqrt{(s+m)!(s-m)!/(2s-1)!}$
is assumed.
The half integer indices $s$ bring some subtlety,
but in all cases the symmetry properties of  $\tilde{f}$
can be written as
\begin{equation} %----
\label{z1}
\tilde{f}_{s_{1}n_{1},s_{2}n_{2}}=-(-1)^{2(m_1m_2-s_1s_2+s_1+s_2)}
\tilde{f}_{s_{2}n_{2},s_{1}n_{1}}.
\end{equation} %---
Now we are going to show that analogues structure takes place for
$N$-point
tree correlation functions for arbitrary  $N>3$.  In $s$-channel  we have
\begin {eqnarray} %-------------------------------------
\label {13}
{\cal
A}_{N}((s_{1},n_{1})^{+}(s_{2},n_{2})^{+}...(s_{N-1},n_{N-1})^{+}
(s_{N},n_{N})^{-})=\int _{0}^{1} dz_{1}\int _{0}^{z_{1}} dz_{2}...\\
\int _{0}^{z_{N-4}} dz_{N-3}<Y^{+}_{s_{1},n_{1}}(\infty )
Y^{+}_{s_{2},n_{2}}(1)W^{+}_{s_{3},n_{3}}(z_{1})
...W^{+}_{s_{N-1},n_{N-1}}(z_{N-3})Y^{-}_{s_{N},n_{N}}(0)>
\\ \nonumber
\end   {eqnarray} %---------------------------------------------------
Performing the OPE as in (\ref {6'}) we get
$$
{\cal A}^{l.s.}_{N}((s_{1},n_{1})^{+}(s_{2},n_{2})^{+}...(s_{N-1},n_{N-1})^{+}
(s_{N},n_{N})^{-})=
$$

$$\int _{0}^{1} \frac{dz_{1}}{z_{1}}\int _{0}^{z_{1}}\frac{dz_{2}}{z_{2}} ...
\int _{0}^{z_{N-4}}\frac{dz_{N-3}}{z_{N-3}}
\sum _{s^{(1)},n^{(1)}}...
\sum _{s^{(N-3)}n^{(N-3)}}
<Y^{+}_{s_{1},n_{1}}
Y^{+}_{s_{2},n_{2}}Y^{-}_{s^{(1)},-n^{(1)}}>$$

\begin{equation}%-------------------------------------------
\label{z2}
<Y^{+}_{s^{(1)},n^{(1)}}Y^{+}_{s_{3},n_{3}}
Y^{-}_{s^{(2)},-n^{(2)}}>
...
<Y^{+}_{s^{(N-3)},n^{(N-3)}}Y^{+}_{s_{N-1},n_{N-1}}Y^{-}_{s_{N},n_{N}}>.
 \end   {equation} %---------------------------------------------------
  So under the analytical  prescription (\ref {6}) we get the following answer
\begin {equation} %---------------------------------------------------
                                                          \label {18}
{\cal A}^{l.s.}_{N}((s_{1},n_{1})^{+}(s_{2},n_{2})^{+}...(s_{N-1},n_{N-1})^{+}
(s_{N},n_{N})^{-})=\frac{1}{\epsilon ^{N-3}} \tilde{{\cal A}}.
\end{equation}
Therefore the $N$-point amplitude for $(N-1)$ (+) states and one (-) state
after renormalization can be represented as \begin {equation}
\label {19}
\tilde
{\cal A}(N-1,1)= \sum _{s^{(1)},n^{(1)}}...\sum _{s^{(N-3)},n^{(N-3)}}
f_{s_{1}n_{1},s_{2}n_{2}}^{s^{(1)}-n^{(1)}}f_{s^{(1)}n^{(1)},
s_{3}n_{3}}^{s^{(2)}-n^{(2)} }...f_{s^{(N-3)}n^{(N-3)},s_{N-1}n_{N-1}}^
{s_{N}n_{N}}.
\end{equation} %---------------------------------------------------
One has a factorisation of the $N$-point amplitude in terms of 3-point one.
The  divergences of discrete states amplitudes can be removed by
renormalization of
wave function and coupling constant  in the string field
theory approach
\begin {equation} %---------------------------------------------------
                                                          \label {r}
\Phi \to {\cal Z}_{1}\Phi,~~{\cal Z}_{1}=\epsilon ^{1/2},
\end   {equation} %---------------------------------------------------
where $\Phi$ is a string field,
\begin {equation} %---------------------------------------------------
g\to g {\cal Z}_{0},~~{\cal Z}_{0}=\epsilon ^{-3/2},
     \label {r'}
\end   {equation} %---------------------------------------------------
We perform the momentum-independent
renormalization.  It is interesting to note
that under this renormalization the tachyon correlation functions
 with non-exceptional momentum (non-compactified correlation
functions) vanish and only correlation functions for discrete momentum $T_{n
\sqrt {2}}$  survey. Our renormalization
for tachyon fields  with exceptional momentum
coincides with  momentum-dependent renormalization  \cite {GrossK}
up to some normalization factors.
%***************************************************************
\section   {Renormalized Ward identities and effective Lagrangian}
%*********************************************************************
 From the currents $W$ one can form the charges \cite{WGR}
\begin {equation} %---------------------------------------------------
                                                          \label {w1}
Q^{\pm}_{s,m}=
\oint \frac{dz}{2\pi i} W^{\pm}_{s,m}(z),
\end   {equation} %---------------------------------------------------
These charges form an algebra
\begin {equation} %---------------------------------------------------
                                                          \label {w2}
[Q^{+}_{s_{1},m_{1}},Q^{+}_{s_{2},m_{2}}]_{\pm}=
\tilde{f}_{s_1m_1,s_2m_2}
Q^{+}_{s_{1}+s_{2}-1,m_{1}+m_{2}}
\end   {equation} %---------------------------------------------------
 where $\pm$  is in accordance with (\ref {z1}). This symmetry
was observed in the  context of matrix models \cite {SMM}.

The current conservation $\bar{\p~}W=0$ implies the symmetry Ward identities
on correlation functions
\begin {equation} %---------------------------------------------------
                                                          \label {w3}
\int <Q^{\pm}_{s,m}Y^{+}_{s_{1},m_{1}}...Y^{-}_{s_{N},m_{N}}
\prod\oint b>=0,
\end   {equation} %---------------------------------------------------
where three of vertex operators $Y^{\pm}$ are fixed and the rest integrated.
The OPE's
$W^{+}_{s,m}$ and $Y^{\pm}_{s_1,m_1}$
assumes the formal Ward identities
$$\tilde{f}_{sm,s_1m_1}
\int <Y^{+}_{s_{1}+s-1,m_{1}+m}
Y^{+}_{s_{2},m_{2}}...Y^{-}_{s_N,m_N}\prod\oint b>
\pm $$
$$\tilde{f}_{sm,s_2m_2}
\int <Y^{+}_{s_{1},m_{1}}Y^{+}_{s_{2}+s-1,m_{2}+m}...Y^{-}_{s_N,m_N}\prod\oint
b
>\pm ...$$
\begin {equation} %---------------------------------------------------
                                                          \label {w5}
\pm \tilde{f}_{s-s_N+1~-m-m_N,s_Nm_N}
\int <Y^{+}_{s_{1},m_{1}}Y^{+}_{s_{2},m_{2}}...Y^{-}_{s_N-s+1,m_N+m}\prod\oint
b
>=0
\end   {equation} %---------------------------------------------------
However all correlation functions in (\ref {w5}) are ill-defined. After
renormalization
(\ref {r}) we left with the correlation functions (\ref {19}) which obvious
satisfy the tree-level Ward identities
$$
\tilde{f}_{sm,s_1m_1}
\tilde {\cal A}((s_{1}+s,m_{1}+m)^{+}(s_{2},m_{2})^{+},...
(s_N,m_N)^{-})\pm
$$
$$
\tilde{f}_{sm,s_2m_2}
\tilde {\cal A}((s_{1},m_{1})^{+}(s_{2}+s,m_{2}+m)^{+},...
(s_N,m_N)^{-})\pm ... \pm $$
\begin {equation} %---------------------------------------------------
                                                          \label {w6}
\tilde{f}_{s-s_N+1~-m-m_N,s_Nm_N}
 \tilde {\cal A}((s_{1},m_{1})^{+}(s_{2},m_{2})^{+},...
(s_N-s+1,m_N+m)^{-})=0.
\end   {equation} %---------------------------------------------------
So far we have found that Ward identities are preserved under
the renormalization.

There are additional subtleties in the closed string related with that one
expect the symmetry charges to alter the particle number  \cite {WZ,V,K}.

One can write an effective Lagrangian which
generates  the renormalized tree amplitudes (\ref{19}). This Lagrangian
describes an interaction of two independent fields $\Phi _{s,n}$ and
$\bar{\Phi} _{s,n}$ with indices $s\geq 0,~ -s\leq n\leq s$. The Lagrangian
has the form
\begin {equation} %---------------------------------------------------
                                                          \label {21}
{\cal L}(\Phi ,\bar{\Phi }) =\sum _{a}\bar{\Phi} _{a}\Phi ^{a}
+g\sum _{a,b,c}\Phi ^{a}\Phi ^{b}\bar{\Phi }_{c}f^{~~c}_{ab},
\end   {equation} %---------------------------------------------------
where $\Phi ^{a}= \Phi _{s,n}$ and $\bar{\Phi }_{a}= \bar{\Phi} _{s,-n}$
 and
\begin {equation} %---------------------------------------------------
                                                          \label {z22}
f_{ab}^{~~c}=<Y^{+}_{a}Y^{+}_{b}Y^{-}_{c}>
\end   {equation} %---------------------------------------------------
The fields $\Phi _{s,n}$ and $\bar{\Phi} _{s,n}$ for integer $s$ anticommute
and for half integer $s$ their statistic is in accordance with symmetry
properties (\ref {z1}).
It is evident that the correlation functions (\ref {19}) come from
the Lagrangian (\ref {21}).
The Lagrangian is invariant under infinite number of
infinitesimal transformations

\begin {equation} %---------------------------------------------------
                                                          \label {24}
\delta _{c}\Phi ^{a}= f ^{~~a}_{cb}\Phi ^{b},~~
\delta _{c} \bar{\Phi }_{a}= f _{ac}^{~~b}\bar{\Phi }_{b},
\end   {equation} %---------------------------------------------------

 From this invariance immediately   follow the Wards identities for
correlation functions (\ref{19}).

\section   {Discussion and conclusion}
It is interesting to note that the interaction part of the
Lagrangian (\ref {21}) is nothing but the first three-linear term of generator
$V(\eta ,\frac{\partial}{\partial \eta})$ of the homotopy Lie algebra.
The homotopy Lie algebra  \cite {WZ} is described by the operator
$$            V= f ^{~~a}_{cb}\eta ^{c}\eta ^{b}\frac{\partial }{\eta ^{a}} +
f ^{~~a}_{cbd}\eta ^{c}\eta ^{b}\eta ^{d}\frac{\partial }{\eta ^{a}} +...,
$$
which satisfies the condition$$\{V,V\}=0.$$
We assume the  identification
$$\eta ^{a}\sim \Phi _{a},~\frac{\partial }{\partial \eta ^{a}}\sim
\bar{\Phi }^{a},$$
so the contractions are identical.
This term is also the BRST charge for the group transformations (\ref {24}).

We expect that leading singularities of N-point discrete states loop amplitudes
coming from open string field theory  \cite {AZ} also reduce to the Lagrangian
(\ref {21}) with full coupling constant
$$f_{ab}^{~~c}=\sum _{loop}<Y^{+}_{a}Y^{+}_{b}Y^{-}_{c}>_{loop}
.$$
It is well known that one cannot limited by self-interacting open string and
one
has to introduce an open-closed transition. One can expect that for closed
string due to  \cite {CSFT} one  has to introduces the Lagrangian
\begin {equation} %---------------------------------------------------
                                                          \label {22}
{\cal L}_{closed} (\Theta ,
\bar{\Theta }) =\sum _{a}\bar{\Theta _{a}}\Theta ^{a}
+g\sum _{a,b,c}\Theta ^{a}\Theta ^{b}\bar{\Theta }_{c}f_{ab}^{~~c}+
g^{2}\sum _{a,b,c,d}\Theta ^{a}\Theta ^{b}\Theta ^{c}\bar{\Theta }_{d}
f_{abc}^{~~~d}+...~,~
\end{equation} %---------------------------------------------------
where
\begin {equation} %---------------------------------------------------
                                                          \label {23}
f_{ab...c}^{~~~~~d}=<O_aO_b...O_c\bar{O}^d>.
\end   {equation} %---------------------------------------------------
The interaction is nothing but the  full Witten Zwiebach operator $V$.
One can also say that
the interaction part of the action (\ref {23}) is the  BRST charge
for a non-closed algebra. The action is invariant under
transformations
$$
\delta _c\Theta ^a=
 f_{cb}^{~~a}\Theta ^b +f^{~~~a}_{cbd}\Theta ^a\Theta ^d+...~,
 $$
\begin {equation} %---------------------------------------------------
                                                       \label {26}
 \delta _c\bar{\Theta }_a
=f^{~~b}_{ac}\bar{\Theta}_b +f^{~~~b}_{acd}\Theta
^{d}\bar{\Theta} _b+...~,
\end   {equation}
if we assume that formfactors are
the subject of relations which follow from the nilpotency condition
$\{V,V\}=0$.

It seems that  discrete states in 2D gravity remind resonances in quantum
theory. If one takes into account a perturbation of the model
then for scattering amplitudes one would obtain
an analogue of the Breit-Wiegner formula  \cite {LL}. Note also that for
analogues problem of small denominators in classical mechanics
one should consider non-quadratic terms which drastically change the
behaviour of trajectories  \cite {Arn}. Probably similar
non-perturbative effects will play an important role in 2D string theory and
we expect that a more suitable framework to describe them is string field
theory. One can expect that the model exhibits a phase transition. One of
phases is a topological one with scattering amplitudes described by the
effective action (\ref {21}).

In summary, we have argued that under some renormalization procedure
the scattering amplitudes of 2D open string  discrete states
  reduce to those of the model with the Lagrangian
(\ref{21}).
$$~$$
{\bf ACKNOWLEDGMENT}
$$~$$
The authors are grateful to P.B.Medvedev and I.V.Volovich for useful
discussions.
$$~$$

\end{document}